\documentclass[12pt]{JHEP3}

\pdfoutput = 1
\usepackage{epsfig}
\usepackage{color}
\epsfclipon
\usepackage{multicol}
\usepackage{epsfig,bm}
\usepackage{amssymb,amsmath}

\newcommand{\roughly}[1]{\mathrel{\raise.3ex\hbox{$#1$\kern-0.85em
\lower1ex\hbox{$\sim$}}}}

\def\be{\begin{equation}}
\def\ee{\end{equation}}
\def\ba{\begin{eqnarray}}
\def\ea{\end{eqnarray}}

\def\tepsilon{{\tilde \epsilon}}
\def\ts{{\tilde s}}
\def\teta{{\tilde \eta}}
\def\tlambda{{\tilde \lambda}}
\def\trho{{\tilde \rho}}
\def\talpha{{\tilde \alpha}}
\def\tN{{\tilde N}}
\def\tF{{\tilde F}}
\def\tG{{\tilde G}}
\def\tmu{{\tilde \mu}}
\def\bbepsilon{{\bar \epsilon}}
\def\bbs{{\bar s}}
\def\bbeta{{\bar \eta}}
\def\bblambda{{\bar \lambda}}
\def\bbrho{{\bar \rho}}
\def\bbalpha{{\bar \alpha}}
\def\bbN{{\bar N}}
\def\bbF{{\bar F}}
\def\bbG{{\bar G}}
\def\bbmu{{\bar\mu}}

\usepackage{ulem}

\title{Horizon-preserving dualities and perturbations in non-canonical scalar field cosmologies}

\author{Ghazal Geshnizjani${}^{1}$, William H. Kinney, ${}^{1,2}$, Azadeh Moradinezhad Dizgah${}^{1}$\\
$^1$Department of Physics, University at Buffalo, SUNY, Buffalo, NY, United States of America\\
$^2$Kavli Insitute for Cosmological Physics, University of Chicago, Chicago, IL, United States of America
 }

\date{\today}

\abstract{We generalize the cosmological duality between inflation and cyclic contraction under the interchange $a \leftrightarrow H$ to the case of non-canonical scalar field theories with varying speed of sound. The single duality in the canonical case generalizes to a family of three dualities constructed to leave the cosmological acoustic horizon invariant. We find three classes of models: (I) DBI inflation, (II) the non-canonical generalization of cyclic contraction, and (III) a new cosmological solution with rapidly decreasing speed of sound and relatively slowly growing scale factor, which we dub {\it stalled} cosmology. We construct dual analogs to the inflationary slow roll approximation, and solve for the curvature perturbation in all three cases.  Both cyclic contraction and stalled cosmology predict a strongly blue spectrum for the curvature perturbations inconsistent with observations.}

\preprint{}

\keywords{}

\begin{document}

\section{Introduction}
\label{Introduction}

A question which has received considerable attention in the recent literature is: what is the most general possible dynamics which can explain the observed form of primordial perturbations in the universe? The most widely known example is cosmological inflation \cite{Guth:1980zm,Linde:1981mu,Albrecht:1982wi}, which creates a Gaussian, adiabatic, nearly scale-invariant spectrum of curvature perturbations via a phase of accelerated expansion in the early universe. Alternatives that have been put forward range from contracting scenarios where the universe needs to go through a singular or non-singular bounce (see \cite{Khoury:2011ii} and references there), string gas cosmology, in which the universe is initially static \cite{Nayeri:2005ck}, varying fundamental speed of light \cite{Albrecht:1998ir}, or a rapidly varying speed of sound \cite{ArmendarizPicon:2003ht,ArmendarizPicon:2006if,Magueijo:2008pm, Bessada:2009ns, Magueijo:2009zp}.

The inflationary mechanism for generating superhorizon density perturbations relies on a shrinking Hubble length, measured in comoving units,
\be
R_H = \left(a H\right)^{-1}.
\ee
The Hubble length is commonly referred to as the cosmological ``horizon'', and we adopt this terminology, although $R_H$ corresponds only to an apparent, not a global, horizon. When the comoving Hubble horizon is shrinking, vacuum quantum modes are stretched to superhorizon scales, where they ``freeze out'' and become classical perturbations \cite{Hawking:1982my,Starobinsky:1982ee,Guth:1982ec,Bardeen:1983qw}. In inflation, the Hubble parameter is nearly constant and the scale factor increases approximately exponentially, so that $R_H$ decreases approximately exponentially. However, accelerating expansion is not the only way to achieve a shrinking $R_H$. A simple duality consisting of an exchange $a \leftrightarrow H$ maps inflationary expansion onto quasi-static contraction \cite{Boyle:2004gv,Piao:2004uq,Lidsey:2004xd,Calcagni:2004wu,Kinney:2010qa} while maintaining an exponentially shrinking $R_H$. We review this duality in Sec. \ref{sec:canonical}. 

Furthermore, in non-canonical scalar field theories with varying speed of sound such as  k-inflation \cite{ArmendarizPicon:1999rj} or DBI inflation \cite{Silverstein:2003hf,Alishahiha:2004eh}, the generation of curvature perturbations is determined not by the Hubble horizon $R_H$, but by the acoustic horizon size \cite{Garriga:1999vw} 
\be
R_s = \frac{c_s}{a H}.
\ee
In models such as DBI inflation, curvature perturbations freeze out at the acoustic horizon, and gravitational wave modes freeze out at the Hubble horizon.\footnote{This behavior does not hold in general: see Refs. \cite{Wands:1998yp,Khoury:2010gw,Piao:2010bi,Joyce:2011ta,Joyce:2011kh,Geshnizjani:2011dk}.} Motivated by this, in Sec. \ref{sec:background}, we generalize the duality $a \leftrightarrow H$ to the case of varying speed of sound, with dualities preserving the form of the acoustic horizon, $a \leftrightarrow H$, $a \leftrightarrow c_s^{-1}$, and $H \leftrightarrow c_s^{-1}$. We identify three mutually dual cosmologies: (I) DBI inflation, (II) the non-canonical generalization of the cyclic model, and (III) a new type of expanding cosmology with slowly varying $a$ and $H$, and exponentially varying $c_S$. We construct dual hierarchies of inflationary flow parameters and show that the inflationary flow equations \cite{Kinney:2002qn,Peiris:2007gz,Kinney:2007ag,Bean:2008ga} are invariant under the dualities. This allows us to construct self-consistent small-parameter duals to the inflationary slow roll approximation.  We show that both the cyclic dual ($a \leftrightarrow H$) and the new duality ($a \leftrightarrow c_s^{-1}$) map the nearly scale-invariant perturbation spectrum in the slow-roll limit to a strongly blue spectrum inconsistent with observation.

\section{Dualities in Canonical Models}
\label{sec:canonical}

In this section we review the horizon-preserving duality between inflation and cyclic cosmology \cite{Boyle:2004gv,Piao:2004uq,Lidsey:2004xd,Kinney:2010qa}. This will form the basis of the generalization to non-canonical models in Sec. \ref{sec:background}.

Consider a flat FRW space with energy density dominated by a single scalar field $\phi$ with Lagrangian
\be
{\mathcal L} = \frac{1}{2} g^{\mu\nu} \partial_\mu \phi \partial_\nu \phi - V\left(\phi\right).
\ee
The equations of motion for the spacetime are given by the Friedmann  Equations, which for a homogeneous field $\phi$ are:
\ba
\label{eq:Friedmann}
H^2 = \left(\frac{\dot a}{a}\right)^2 &=& \frac{1}{3 M_P^2}\left[\frac{1}{2} \dot\phi^2 + V\left(\phi\right)\right],\cr
\left(\frac{\ddot a}{a}\right) &=& - \frac{1}{3M_P^2} \left[\dot\phi^2 - V\left(\phi\right)\right]. 
\ea
The equation of motion for the field $\phi$ is
\be
\label{eq:phieom}
\ddot \phi + 3 H \dot\phi + V'\left(\phi\right) = 0.
\ee
If the field evolution is monotonic in time, we can write the scale factor $a\left(\phi\right)$ and Hubble parameter $H\left(\phi\right)$ as functions of the field $\phi$ rather than time, {\it i.e.} we define all of our physical parameters along the trajectory in phase space ${\dot\phi}\left(\phi\right)$ corresponding to the classical solution to the equations of motion. Equations (\ref{eq:Friedmann}) and (\ref{eq:phieom}) can then be re-written exactly in the Hamilton-Jacobi form
\ba
\label{eq:Hamjacobi}
V\left(\phi\right) &=& 3 M_P^2 H^2\left(\phi\right) \left[1 - \frac{2 M_P^2}{3} \left(\frac{H'\left(\phi\right)}{H\left(\phi\right)}\right)^2\right],\cr
\dot\phi &=& - 2 M_P^2 H'\left(\phi\right).
\ea 
Note that if the trajectory of the field in phase space ${\dot\phi}\left(\phi\right)$ is determined, then the potential $V\left(\phi\right)$ similarly fully determined, and the Lagrangian corresponding to a particular classical solution for $\phi$ is unique. This will not be true in the non-canonical generalization.
We define an infinite hierarchy of dimensionless parameters \cite{Copeland:1993jj,Liddle:1994dx} by taking successive derivatives of the Hubble parameter $H$ with respect to the field $\phi$:
\ba
\label{eq:definflationparamshierarchy}
\epsilon &\equiv& {2 M_{P}^2 } \left(\frac{H'(\phi)}{H(\phi)}\right)^2, \cr 
\eta &\equiv& {2 M_{P}^2 } \frac{H''(\phi)}{H(\phi)}, \cr
{}^2 \lambda  &\equiv& {4 M_{P}^4}  \frac{H'(\phi) H'''(\phi)}{H^2(\phi)}, \cr 
\vdots \cr   
{}^ \ell \lambda &\equiv& {\left(2 M_{P}^2\right) ^ \ell} \frac{H'(\phi)^{\left(\ell-1\right)}}{H(\phi)^ \ell} \frac{d^{\left(\ell +1\right)} H(\phi)}{d \phi ^ {\left(\ell +1\right)}},
\ea
where the prime denotes derivatives with respect to scalar field $\phi$. We refer to this hierarchy as {\it flow parameters}. The first flow parameter $\epsilon$ is related to the equation of state by
\be
\epsilon = \frac{3}{2} \left(1 + w\right) = -\frac{a}{H} \frac{d H}{d a}.
\ee
In an expanding universe, the comoving horizon shrinks for $w < -1/3$, or $\epsilon < 1$, which is just the case of inflation. For a canonical scalar field, the null energy condition $\rho + p \geq 0$ is satisfied and $\epsilon \geq 0$. In an inflating universe, the scale factor increases quasi-exponentially, with the Hubble parameter $H \simeq {\rm const.}$ and
\be
a \propto \exp\left[\int{H dt}\right] \equiv e^{-N},
\ee
where $N$ is the number of e-folds. The sign convention for $N$ above means that $N$ {\it decreases} with increasing time, and therefore measures the number of e-folds before the end of inflation. The number of e-folds $N$ is related to the field $\phi$ by
\ba
\label{numberofefolds}
d N \equiv - H dt &=& -\frac{d a(\phi)}{a(\phi)} \cr
 &=& \frac{1}{\sqrt{2} M_P}\frac{d\phi}{\sqrt{\epsilon(\phi)}},
\ea
where we choose the sign convention that $\sqrt{\epsilon}$ has the same sign as $H'\left(\phi\right)$,
\be
\sqrt{\epsilon} \equiv + \sqrt{2} M_P \frac{H'\left(\phi\right)}{H\left(\phi\right)}.
\ee
We can then write the parameter $\epsilon$ as
\be
\epsilon = \frac{1}{H} \frac{d H}{d N}.
\ee
Similarly, taking derivatives of the flow parameters with respect to $N$, we can generate an infinite set of differential equations relating the parameters:
\ba 
\frac{d \epsilon}{d N} &=& 2 \epsilon \left(\eta - \epsilon\right), \cr
\frac{d \eta}{d N} &=&  {}^2 \lambda - \epsilon \eta, \cr
\vdots \cr  
\frac{d {}^\ell \lambda} {d N} &=& \left[(\ell-1) \eta - \ell \epsilon \right] {}^ \ell \lambda + {} ^ {(\ell + 1)} \lambda.
\ea
Taken to infinite order, this hierarchy of flow equations completely specifies the evolution of the spacetime \cite{Kinney:2002qn}: once the parameters $\epsilon\left(N\right)$, $\eta\left(N\right)$ and so forth are known,\footnote{If a lower order parameter is known higher order parameters can be derived from it. For example, $\eta(N)$ can be derived from $\epsilon(N)$.} the cosmological dynamics and scalar field potential are fixed via the Hamilton-Jacobi Equations (\ref{eq:Hamjacobi}). 

We wish to construct a similar hierarchy of flow equations applicable to contracting universes, utilizing the duality $a \leftrightarrow H$, which preserves the behavior of the comoving Hubble horizon. Under the duality $a \leftrightarrow H$, the parameter $\epsilon$ transforms as
\be
\epsilon = 2 M_P^2 \left(\frac{H'\left(\phi\right)}{H\left(\phi\right)}\right)^2 \rightarrow {\tilde \epsilon} = 2 M_P^2 \left(\frac{a'\left(\phi\right)}{a\left(\phi\right)}\right)^2.
\ee
From the definition of Hubble parameter we have
\be
\frac{H'(\phi)}{H(\phi)} = -\frac{1}{2 M_P^2} \frac{a(\phi)}{a'(\phi)},
\ee
so that the dual parameter $\tilde \epsilon$ can be seen to be the inverse of the flow parameter $\epsilon$,
\be
\label{eq:epsinv}
{\tilde \epsilon} = \frac{1}{2 M_P^2} \left(\frac{H(\phi)}{H'(\phi)}\right)^2 = \frac{1}{\epsilon}.
\ee
Under $a \leftrightarrow H$, the dual of the number of e-folds $d N = - da / a$ is
\ba
\label{eq:dualN}
d \tilde{N} \equiv - \frac{d H(\phi)}{H(\phi)} &=& -\frac{H'(\phi)}{H(\phi)}d\phi \cr
 &=& \frac{1}{\sqrt{2} M_P} \frac{d \phi}{\sqrt{{\tilde \epsilon}(\phi)}},
\ea
where we choose the sign convention that $\sqrt{\tilde \epsilon}$ has the same sign as $a'\left(\phi\right)$ and therefore the {\it opposite} sign as $H'\left(\phi\right) / H\left(\phi\right)$,
\be
\sqrt{\tilde \epsilon} \equiv + \sqrt{2} M_P \frac{a'\left(\phi\right)}{a\left(\phi\right)} = -\frac{1}{2 M_p^2} \frac{H\left(\phi\right)}{H'\left(\phi\right)}.
\ee
The number of e-folds and its dual are related by
\be
d N = - {\tilde\epsilon} d \tilde{N},
\ee
so that we can write $\epsilon$ and ${\tilde \epsilon}$ as
\be
\epsilon = - \frac{d {\tilde N}}{d N},\qquad \qquad {\tilde \epsilon} = - \frac{d N}{d {\tilde N}}.
\ee
Defining ${\mathcal N}$ as the logarithm of the comoving horizon size,
\be
{\mathcal N} \equiv - \ln{\left(a H\right)},
\ee
we see that ${\mathcal N}$ is invariant under the duality $a \leftrightarrow H$, and is related to $N$ and ${\tilde N}$ by
\be
d {\mathcal N} = \left(1 - \epsilon\right) dN = \left(1 - {\tilde \epsilon}\right) d {\tilde N},
\ee
In an inflating universe, the comoving horizon shrinks ($d {\mathcal N} / dN > 0$) for $\epsilon < 1$. Likewise, in a contracting universe, the comoving horizon shrinks for $\tilde \epsilon < 1$, so that $\tilde\epsilon$ plays the same role in contracting universes that $\epsilon$ plays in expanding universes. We define the full hierarchy of dual flow parameters by replacing $H\left(\phi\right)$ in Eqs.(\ref{eq:definflationparamshierarchy}) with $a\left(\phi\right)$,  
\ba
\label{eq:defcontractingparamshierarchy}
{\tilde\epsilon} &\equiv& {2 M_{P}^2 } \left(\frac{a'(\phi)}{a(\phi)}\right)^2, \cr 
\tilde{\eta} &\equiv& {2 M_{P}^2 } \frac{a''(\phi)}{a(\phi)}, \cr
 {}^2 \tilde{\lambda}  &\equiv& {4 M_{P}^4}  \frac{a'(\phi) a'''(\phi)}{a^2(\phi)}, \cr 
\vdots \cr   
{}^ \ell \tilde{\lambda} &\equiv& {\left(2 M_{P}^2\right) ^ \ell} \frac{a'(\phi)^{\left(\ell-1\right)}}{a(\phi)^ \ell} \frac{d^{\left(\ell +1\right)} a(\phi)}{d \phi ^ {\left(\ell +1\right)}},
\ea
Comparing Eqs. (\ref{numberofefolds}) and (\ref{eq:dualN}), we see that the dual $d {\tilde N}$ has exactly the same form as $d N$, so that the dual flow equations are identical to their counterparts in inflation:
\ba
\label{contractingflow} 
\frac{d {\tilde\epsilon}}{d \tilde{N}} &=& 2 {\tilde\epsilon} \left(\tilde{\eta} - {\tilde\epsilon}\right), \cr
\frac{d \tilde{\eta}}{d \tilde{N}} &=& {}^2\tilde{\lambda}-{\tilde\epsilon}\tilde{\eta}, \cr
\vdots \cr  
\frac{d {}^\ell \tilde{\lambda}}{d \tilde{N}} &=& \left[(\ell-1) \tilde{\eta} - \ell {\tilde\epsilon}\right] {}^ \ell \tilde{\lambda} + {} ^ {(\ell + 1)}\tilde{\lambda}.
\ea
This duality invariance of the flow equations means that for every inflationary potential, there exists an equivalent dual cyclic potential for which the comoving Hubble horizon has the same dynamics \cite{Boyle:2004gv}.

We can then define a dual analog of the slow-roll limit of inflation by taking ${\tilde \epsilon} \ll 1$ and ${\tilde \eta} \ll 1$, so that 
\be
\frac{d {\tilde \epsilon}}{d {\tilde N}} \ll {\tilde \epsilon},
\ee
and so forth, with higher order flow parameters small. The flow parameters can then be expressed approximately in terms of the potential by using the Hamilton-Jacobi Equations,
\ba
\label{eq:Hamjacobidual}
V\left(\phi\right) &=& 3 M_P^2 H^2\left(\phi\right) \left(1 - \frac{1}{3 {\tilde \epsilon}\left(\phi\right)}\right), \cr
\dot\phi &=& - 2 M_P^2 H'\left(\phi\right).
\ea
For the dual-slow roll limit $\tilde\epsilon \ll 1$,
\ba
\label{eq:VdualSR}
V\left(\phi\right) &\simeq& - M_P^2 \frac{H^2\left(\phi\right)}{{\tilde \epsilon}\left(\phi\right)}\cr
&=& - 2 M_P^4 \left[H'\left(\phi\right)\right]^2 = -\frac{1}{2} \dot\phi^2.
\ea
In the case of slow roll inflation, $\epsilon \propto \left(H'/H\right)^2 \ll 1$ means that the Hubble parameter $H$ and scale factor $a$ evolve as
\ba
H^2 &\simeq& \frac{1}{3 M_P^2} V\left(\phi\right) \simeq {\rm const.},\cr
a &\sim& e^{H t}.
\ea
In the dual limit ${\tilde \epsilon} \propto \left(a'/a\right)^2 \ll 1$, we have
\ba
a &\simeq& {\rm const.},\cr
\left(\frac{H}{M_p}\right)^2 &=& \frac{1}{3 M_P^4} \left[\frac{1}{2}\dot\phi^2 + V\left(\phi\right)\right] \simeq 0.
\ea
The dual limit of slow roll inflation is therefore a quasi-static contraction, with the scale factor nearly time-invariant. (Here, and in what follows, we use the term ``quasi-static'' to indicate that the scale factor $a$ is varying slowly in comparison to other physical parameters such as $H$ or $\gamma$.) Note that the potential in the cyclic dual to inflation must be negative.  Differentiating Eq. (\ref{eq:VdualSR}) with respect to $\phi$ gives
\ba
\frac{V'\left(\phi\right)}{V\left(\phi\right)} &\simeq& 2 \frac{H'}{H} - \frac{{\tilde \epsilon}'}{\tilde \epsilon}\cr
&=& 2 \frac{H'}{H} \left(1 + {\tilde \eta} - {\tilde \epsilon}\right),
\ea
where we have used the flow equations (\ref{contractingflow}) and
\be
\frac{d}{d \phi} = \frac{1}{M_P \sqrt{2 {\tilde \epsilon}}} \frac{d}{d {\tilde N}} = - \frac{H'}{H} \frac{d} {d {\tilde N}}.
\ee
Then, to lowest order in the flow parameters,
\be
\tilde\epsilon = \frac{1}{2 M_P^2} \left(\frac{H\left(\phi\right)}{H'\left(\phi\right)}\right)^2 \simeq \frac{2}{M_P^2} \left(\frac{V\left(\phi\right)}{V'\left(\phi\right)}\right)^2.
\ee
Using the first flow equation (\ref{contractingflow}), the second flow parameter $\tilde\eta$ can be written in the small parameter limit as
\ba
{\tilde \eta} &=& {\tilde \epsilon} + \frac{1}{2 {\tilde \epsilon}} \frac{d {\tilde \epsilon}}{d {\tilde N}}\cr
&\simeq& -2 \left[1 - \frac{1}{M_P^2} \left(\frac{V}{V'}\right)^2 - \frac{V V''}{\left(V'\right)^2}\right].
\ea
This differs slightly from the ``fast roll'' parameter $\eta$ defined in Ref. \cite{Khoury:2003rt}, but is physically equivalent. We hesitate to adopt the terminology ``fast roll'' because the dual limit of slow roll inflation is not the limit of $\dot\phi^2 \gg V\left(\phi\right)$, but the highly tuned circumstance that the kinetic and potential contributions to the energy density almost exactly cancel. We will use the term {\it dual-slow roll} to describe this situation. 

\section{Generalized Dualities}
\label{sec:background}

In this section we consider the case of cosmologies with a time-varying speed of sound. In this case, the relevant horizon for mode generation is not the Hubble horizon, but the sound horizon, given by:
\be
R_s = \frac{c_S}{a H} = \frac{1}{\gamma a H},
\ee
where $\gamma \equiv c_s^{-1}$. This suggests a natural generalization of the horizon-preserving duality in Sec. \ref{sec:canonical}: in addition to the duality under the interchange $a \leftrightarrow H$, we can construct additional dualities under the interchanges:
\ba
a &\leftrightarrow& \gamma,\cr
H &\leftrightarrow& \gamma.
\ea
 
We begin by deriving a fully general version of the Hamilton-Jacobi equations from Sec. \ref{sec:canonical}, applicable to non-canonical Lagrangians \cite{Bean:2008ga}. Consider a general Lagrangian ${\mathcal L}\left(X,\phi\right)$, where in the homogeneous limit,
\be
\label{eq:defkinetic}
X \equiv \frac{1}{2} g^{\mu\nu} \partial_\mu \phi \partial_\nu \phi = \frac{1}{2} \dot\phi^2.
\ee
The energy density and pressure are then
\ba
&&p = {\mathcal L},\cr
&&\rho = 2 X {\mathcal L}_X - {\mathcal L},
\ea
where subscripts denote a derivative,
\be
{\mathcal L}_X \equiv \frac{\partial {\mathcal L}}{\partial X}.
\ee
The speed of sound $c_s$ is given by
\be
\label{eq:generalcs}
c_s^2 \equiv \frac{p_X}{\rho_X} = \frac{{\mathcal L}_{X}}{{\mathcal L}_{X} + 2 X {\mathcal L}_{XX}}.
\ee
The Friedmann equation can then be written
\be
\label{eq:generalFriedmann}
3 M_P^2 H^2 = 2 X {\mathcal L}_X - {\mathcal L},
\ee
and the continuity equation ${\dot \rho} + 3 H \left(\rho + p\right) = 0$ becomes
\be
\label{eq:generalcontinuity}
M_P^2 {\dot H} = - X {\mathcal L}_X.
\ee
We construct a general form of the Hamilton-Jacobi equations by taking the field evolution to be monotonic in $\phi$, so the kinetic term is described by a trajectory $\sqrt{2 X} = \dot\phi\left(\phi\right)$. Then the continuity equation gives
\be
M_P^2 {\dot H} = M_P^2 H'\left(\phi\right) {\dot \phi} = - \frac{1}{2} {\dot\phi}^2 {\mathcal L}_X,
\ee
so that 
\be
\label{eq:generalphidot}
{\dot\phi} = -\frac{2 M_P^2}{{\mathcal L}_X} H'\left(\phi\right).
\ee
Now substituting for $\dot\phi$ in Eq.(\ref{eq:generalFriedmann}), we have a generalized Hamilton-Jacobi Equation,
\be
\label{eq:generalHamiltonJacobi}
{\mathcal L} = - 3 M_P^2 H^2\left(\phi\right) \left[1 -  \frac{4 M_P^2}{3 {\mathcal L}_X} \left(\frac{H'}{H}\right)^2\right].
\ee
(In what follows, unless otherwise noted, a prime denotes a derivative with respect to the field, {\it i.e.} $H' \equiv H'\left(\phi\right)$.) Note that the non-canonical generalization of the Hamilton-Jacobi equations introduces a new ambiguity when ``reconstructing'' the Lagrangian from a field solution. The Lagrangian corresponding to a particular classical solution $X\left(\phi\right)$, is not unique, and a given trajectory in phase space corresponds to a family of Lagrangians ${\mathcal L}\left(X,\phi\right)$ defined over the full phase space. More than one Lagrangian can have the same solution, and fixing a unique Lagrangian requires an additional arbitrary constraint, or choice of {\it gauge} \cite{Bean:2008ga}.  This gauge freedom is essential to constructing self-consistent duals to DBI inflation.

We proceed as in Sec. \ref{sec:canonical} by defining dual versions of the inflationary flow hierarchy. We will show that there exist three distinct types of cosmological evolution, which we refer to as Type I, Type II, and Type III, which Type I being standard inflationary cosmology, Type II being the generalized cyclic scenario, and Type III being a novel expanding cosmological solution with rapidly varying speed of sound and slowly varying scale factor and Hubble parameter.

\subsection{Type I Cosmology: DBI Inflation}

We begin with the familiar case of DBI Inflation, consisting of accelerating expansion with varying speed of sound. We first derive a useful identity from the definition of the Hubble parameter,
\be
H \equiv \frac{\dot a}{a} = \frac{a'\left(\phi\right)}{a\left(\phi\right)} {\dot\phi},
\ee
which, using Eq. (\ref{eq:generalphidot}), results in the identity
\be
\label{eq:usefulidentity}
{\mathcal L}_X = - 2 M_P^2 \left(\frac{H'}{H}\right) \left(\frac{a'}{a}\right).
\ee
Therefore we can write the Hamilton-Jacobi equation (\ref{eq:generalHamiltonJacobi}) as
\ba
-{\mathcal L} &=& 3 M_P^2 H^2 \left[1 + \frac{2}{3} \left(\frac{a}{H} \frac{d H}{d a}\right)\right]\cr
              &=& 3 M_P^2 H^2 \left[1 - \frac{2}{3} \left(\frac{1}{H} \frac{d H}{d N}\right)\right],
\ea
where we define the number of e-folds as in the case of canonical inflation, 
\be
\label{eq:TypeIN}
dN \equiv - da / a = - H dt. 
\ee
The sign convention for $d N$ is identical to a the canonical case: For $d t > 0$, an expanding universe ($H > 0$) corresponds to $d N < 0$, and a contracting universe corresponds to $d N > 0$. We  define a general form for the first flow parameter $\epsilon$
\be
\label{eq:TypeIepsilon}
\epsilon \equiv \frac{1}{H}\frac{d H}{d N} = \frac{2 M_P^2}{{\mathcal L}_X} \left(\frac{H'\left(\phi\right)}{H\left(\phi\right)}\right)^2 = \frac{3}{2} \left(1 + \frac{p}{\rho}\right).
\ee
We can then write the general Hamilton-Jacobi Equations in terms of $\epsilon\left(\phi\right)$ as:
\ba 
\label{eq:TypeIHamiltonJacobi}
&&{\dot\phi} = -\frac{2 M_P^2}{{\mathcal L}_X} H'\left(\phi\right),\cr
&&-{\mathcal L} = 3 M_P^2 H^2 \left[1 + \frac{2}{3} \epsilon\right],
\ea
We then define a new flow parameter in terms of the speed of sound $c_s$,
\be
\label{eq:TypeIs}
s \equiv - \frac{1}{c_s} \frac{d c_s}{d N} = \frac{1}{\gamma} \frac{d \gamma}{d N} = \frac{2 M_P^2}{{\mathcal L}_X} \frac{H'}{H} \left(\frac{\gamma'}{\gamma}\right).
\ee
With these definitions, we have a set of differential relations between the number of e-folds $N$ and the dynamical parameters $a$, $H$, and $\gamma$ specifying the evolution of the spacetime
\be
\label{eq:TypeIdifferential}
d N = - \frac{da}{a} = \frac{1}{\epsilon} \frac{d H}{H} = \frac{1}{s} \frac{d \gamma}{\gamma}.
\ee
We define ${\mathcal N}$ as the logarithm of the comoving acoustic horizon size,
\be
\label{eq:loghorizon}
{\mathcal N} \equiv - \ln{\left(\gamma a H\right)},
\ee
which can be related to the number of e-folds $N$ as
\be
d {\mathcal N} = \left(1 - \epsilon - s\right) d N.
\ee
Therefore, in an expanding universe, $d N < 0$, the acoustic horizon shrinks ($d {\mathcal N} < 0$) for $1 - \epsilon - s > 0$, and grows for $1 - \epsilon -s < 0$. In a contracting universe, $d N > 0$, the acoustic horizon shrinks for $1 - \epsilon -s < 0$ and grows for $1 - \epsilon - s > 0$. Using the Hamilton-Jacobi Equation (\ref{eq:TypeIHamiltonJacobi}), we can write $N$ in terms of the field $\phi$ as
\be
\label{eq:TypeINphi}
d N = \frac{1}{\epsilon}\frac{H'}{H} d\phi = \sqrt{\frac{{\mathcal L}_X}{2 M_P^2 \epsilon}} d\phi.
\ee
It is straightforward to construct an exact solution for the spacetime evolution by taking $\epsilon = {\rm const.}$ and $s = {\rm const.}$, 
\ba
\label{eq:TypeIpowerlaw}
a &\propto& e^{-N},\cr
H &\propto& e^{\epsilon N},\cr
\gamma &\propto& e^{s N}.
\ea
We can recognize this solution as power-law evolution from the definition of the number of e-folds (\ref{eq:TypeIN}),
\be
dt = - dN / H = e^{-\epsilon N} dN,
\ee
with solution
\be
t \propto e^{-\epsilon N}.
\ee
Then, from (\ref{eq:TypeIpowerlaw}),
\ba
\label{eq:TypeIt}
a(t) &\propto& t^{1 / \epsilon},\cr
H(t) &\propto& t^{-1},\cr
\gamma(t) &\propto& t^{- s / \epsilon}.
\ea
We emphasize that the power-law solution (\ref{eq:TypeIpowerlaw}) is a particular solution to the Friedmann Equation, while the differential relations (\ref{eq:TypeIdifferential}) are general and apply even when $\epsilon$ and $s$ are not constant. In the limit $\epsilon \ll 1$ and $s \ll 1$, the power-law solution corresponds to quasi-static $H$ and $\gamma$, with rapidly varying scale factor, {\it i.e.} inflation. 

To make contact with the associated field Lagrangian, we consider the expressions (\ref{eq:TypeIepsilon}) and (\ref{eq:TypeIs}) for $\epsilon$ and $s$ in terms of the field $\phi$:
\ba
\epsilon &=& \frac{2 M_P^2}{{\mathcal L}_X} \left(\frac{H'\left(\phi\right)}{H\left(\phi\right)}\right)^2 = {\rm const.},\cr
s &=& \frac{2 M_P^2}{{\mathcal L}_X} \frac{H'}{H} \left(\frac{\gamma'}{\gamma}\right) = {\rm const.}
\ea
Note that we have two equations of three free functions, $H\left(\phi\right)$, $\gamma\left(\phi\right)$, and ${\mathcal L}_X\left(\phi\right)$. The extra degree of freedom corresponds to a choice of gauge, and is therefore unphysical \cite{Bean:2008ga}. We can recover the well-studied DBI Lagrangian by taking the following gauge condition to be satisfied for every solution on the phase space for $\phi$ and $\dot{\phi}$: 
\be
\label{eq:DBIgauge}
{\mathcal L}_X = \gamma,
\ee
so that the identity (\ref{eq:usefulidentity}) becomes
\be
\label{eq:TypeIidentity}
\frac{2 M_P^2}{\gamma} \left(\frac{H'}{H}\right) \left(\frac{a'}{a}\right) = -1,
\ee
and
\ba
\label{eq:DBIepsilons}
\epsilon &=& \frac{2 M_P^2}{\gamma\left(\phi\right)} \left(\frac{H'\left(\phi\right)}{H\left(\phi\right)}\right)^2,\cr
s &=&  \frac{2 M_P^2}{\gamma\left(\phi\right)} \left(\frac{\gamma'\left(\phi\right)}{\gamma\left(\phi\right)}\right) \left(\frac{H'\left(\phi\right)}{H\left(\phi\right)}\right).
\ea
For $\epsilon$ and $s$ constant, $H$ and $\gamma$ can be expressed as functions of $\phi$ \cite{Kinney:2007ag},
\ba
\gamma &\propto& \phi^{-2},\cr
H &\propto& \phi^{- 2 \epsilon / s}.
\ea
The gauge condition (\ref{eq:DBIgauge}), combined with the expression (\ref{eq:generalcs}) for the speed of sound, leads to a relation 
\be
\gamma^2 = {\mathcal L}_X^2 = 1 + 2 X \frac{{\mathcal L}_{XX}}{{\mathcal L}_X}.
\ee
This can be re-written as a differential equation for ${\mathcal L}$,
\be
2 X {\mathcal L}_{XX} + {\mathcal L}_X - {\mathcal L}_X^3 = 0,
\ee
with solution 
\be
\label{eq:DBILagrangian}
{\mathcal L} = f^{-1}\left(\phi\right) \sqrt{1 + f\left(\phi\right) X} - V\left(\phi\right),
\ee
where the integration constants $f\left(\phi\right)$ and $V\left(\phi\right)$ are related to $H$ and $\gamma$ by the Hamilton-Jacobi Equations (\ref{eq:TypeIHamiltonJacobi}) \cite{Kinney:2007ag}
\ba
\label{eq:DBIfunctions}
f\left(\phi\right) &=& \frac{1}{2 M_{P}^2 H^2\left(\phi\right)\epsilon}\left(\frac{\gamma^2\left(\phi\right) - 1}{\gamma\left(\phi\right)}\right),\cr
V\left(\phi\right) &=& 3 M_P^2 H^2\left(\phi\right) \left(1 - \frac{2\epsilon}{3} \frac{\gamma\left(\phi\right)}{\gamma\left(\phi\right)+1}\right).
\ea
We have therefore constructed a two-parameter family of exact solutions to the Friedmann Equations giving power-law evolution for $a\left(t\right)$ and $c_s\left(t\right)$ (\ref{eq:TypeIt}), including a fully specified field Lagrangian (\ref{eq:DBILagrangian}),(\ref{eq:DBIfunctions}).

To generalize to the case of time-dependent $\epsilon$ and $s$, we can define an infinite hierarchy of flow parameters \cite{Peiris:2007gz,Kinney:2007ag},
\ba
\label{eq:TypeIflowparams}
\eta\left(\phi\right) &\equiv& \frac{2 M_P^2}{\gamma\left(\phi\right)} \frac{H''\left(\phi\right)}{H\left(\phi\right)},\cr
\rho\left(\phi\right) &\equiv& \frac{2 M_P^2}{\gamma\left(\phi\right)}  \frac{\gamma''\left(\phi\right)}{\gamma\left(\phi\right)},\cr
{}^\ell \lambda\left(\phi\right) &\equiv& \left(\frac{2 M_P^2}{\gamma\left(\phi\right)}\right)^{\ell} \left(\frac{H'\left(\phi\right)}{H\left(\phi\right)}\right)^{\ell - 1} \frac{1}{H\left(\phi\right)} \frac{d^{\ell + 1} H\left(\phi\right)}{d \phi^{\ell + 1}},\cr
{}^\ell \alpha\left(\phi\right) &\equiv& \left(\frac{2 M_P^2}{\gamma\left(\phi\right)}\right)^{\ell} \left(\frac{H'\left(\phi\right)}{H\left(\phi\right)}\right)^{\ell - 1} \frac{1}{\gamma\left(\phi\right)} \frac{d^{\ell + 1} \gamma\left(\phi\right)}{d \phi^{\ell + 1}},
\ea
where $\ell = 2,\ldots,\infty$ is an integer index.\footnote{Our notation differs slightly from that of Peiris {\it et al.} \cite{Peiris:2007gz}, who use $\kappa$ in place of s.} This reduces to the usual canonical flow hierarchy \cite{Kinney:2002qn} for $\gamma = 1$. We can write the number of e-folds (\ref{eq:TypeINphi}) as
\be
\label{eq:DBIN}
d N \equiv - H dt = \frac{\gamma}{2 M_p^2} \left(\frac{H}{H'}\right) d\phi,
\ee
so that the flow parameters (\ref{eq:TypeIflowparams}) are related by a series of first-order {\it flow equations},
\ba
\label{eq:TypeIflowequations}
\epsilon &=& \frac{1}{H}\frac{d H}{d N},\cr
\frac{d \epsilon}{d N} &=& \epsilon\left(2 \eta - 2 \epsilon - s\right),\cr
\frac{d \eta}{d N} &=& -\eta\left(\epsilon + s\right) + {}^2 \lambda,\cr
\frac{d {}^\ell \lambda}{d N} &=& -{}^\ell \lambda \left[\ell \left(s + \epsilon\right) - \left(\ell - 1\right) \eta\right] + {}^{\ell + 1}\lambda,\cr
s &=& \frac{1}{\gamma}\frac{d \gamma}{d N},\cr
\frac{d s}{d N} &=& -s \left(2 s + \epsilon - \eta\right) + \epsilon \rho,\cr
\frac{d \rho}{d N} &=& -2 \rho s + {}^2 \alpha,\cr
\frac{d {}^\ell \alpha}{d N} &=& -{}^\ell \alpha \left[\left(\ell + 1\right) s + \left(\ell - 1\right) (\epsilon-\eta)\right] + {}^{\ell + 1} \alpha.
\ea
A solution to this set of differential equations $\epsilon\left(N\right),\ s\left(N\right),\ \eta\left(N\right),\ \rho\left(N\right),\ \ldots$ completely specifies the evolution of the spacetime, via the functions $H\left[N\left(\phi\right)\right]$ and $\gamma\left[N\left(\phi\right)\right]$. This formulation of the dynamics is useful for numerical investigation of the full non-canonical inflationary parameter space \cite{Powell:2008bi,Easson:2009wc,Kawasaki:2009yn,Easson:2010uw,Easson:2010zy}.

\subsection{Type II Cosmology: the cyclic dual}

In this section we construct the non-canonical generalization of the duality between inflation and cyclic scenarios \cite{Boyle:2004gv,Piao:2004uq,Lidsey:2004xd,Kinney:2010qa} reviewed in Sec. \ref{sec:canonical}. Under this duality, the dynamics of the scale factor and the Hubble parameter are interchanged, {\it i.e.} $a \leftrightarrow H$,
\ba
&&\tepsilon \equiv \frac{1}{a}\frac{da}{d \tN},\cr
&&\ts \equiv \frac{1}{\gamma}\frac{d \gamma}{d \tN},\cr
&&\frac{1}{H}\frac{d H}{d \tN} \equiv -1.
\ea
The dual number of e-folds $\tN$ can be related to the number of e-folds $N$ in a Type I cosmology by
\be
d \tN = \frac{1}{\tepsilon} \frac{da}{a} = - \frac{1}{\tepsilon} d N = \frac{H}{\tepsilon} dt.
\ee
Note that $d t > 0$ and the sign convention $d \tN < 0$ implies $H < 0$, or a contracting cosmology, so that the $a \leftrightarrow H$ duality swaps between expansion and contraction. Using
\be
\label{eq:TypeIIHa}
\frac{d H}{H} = - \frac{1}{\tepsilon}\frac{d a}{a},
\ee
and Eq. (\ref{eq:usefulidentity}), we have
\be
{\mathcal L}_X = \frac{2 M_P^2}{\tepsilon} \left(\frac{a'}{a}\right)^2,
\ee
which gives us a dual expression for the parameter $\tepsilon$ in terms of the field $\phi$,
\be
\label{eq:TypeIIepsilon}
\tepsilon = \frac{2 M_P^2}{{\mathcal L}_X} \left(\frac{a'\left(\phi\right)}{a\left(\phi\right)}\right)^2.
\ee
Similarly, from 
\be
\frac{d H}{H} = - \frac{1}{\ts}\frac{d \gamma}{\gamma},
\ee
we have an expression for $\ts$ in terms of the field $\phi$,
\be
\label{eq:TypeIIs}
\ts =  \frac{2 M_P^2}{{\mathcal L}_X} \left(\frac{\gamma'\left(\phi\right)}{\gamma\left(\phi\right)}\right)\left(\frac{a'\left(\phi\right)}{a\left(\phi\right)}\right).
\ee
Note that these expressions for $\tepsilon$ and $\ts$ can be obtained by directly applying the duality $a \leftrightarrow H$ to the expressions (\ref{eq:TypeIepsilon}) and (\ref{eq:TypeIs}). From Eq. (\ref{eq:TypeIdifferential}) and
\be 
d \tN = \frac{1}{\tepsilon} \frac{da}{a} = -\frac{d H}{H} = \frac{1}{\ts} \frac{d \gamma}{\gamma},
\ee
we can relate the dual parameters to their counterparts in Type I,
\ba
&&\tepsilon = \frac{1}{\epsilon},\cr
&&\ts = - \frac{s}{\epsilon}.
\ea
Since $\epsilon$ is positive-definite, so is $\tepsilon$, and therefore for $d \tN < 0$, we have $da < 0$, consistent with a negative Hubble parameter. The logarithm of the acoustic horizon size $\mathcal N$ can then be written in terms of the number of e-folds $\tN$ as
\be
d {\mathcal N} = \left(1 - \tepsilon - \ts\right) d \tN.
\ee
Therefore, for $d \tN < 0$, the acoustic horizon shrinks when $1 - \tepsilon - \ts < 0$. As in the case of a Type I cosmology, we can write the number of e-folds $\tN$ in terms of the field $\phi$ as:
\be
\label{eq:TypeIINphi}
d \tN = \frac{1}{\tepsilon} \frac{H}{\dot \phi} d\phi = \sqrt{\frac{{\mathcal L}_X \tepsilon}{2 M_P^2}} d \phi,
\ee
which can be seen as the $\epsilon \rightarrow 1 / \tepsilon$ dual of Eq. (\ref{eq:TypeINphi}). We construct an exact solution for the spacetime evolution by taking $\tepsilon = {\rm const.}$ and $\ts = {\rm const.}$, 
\ba
\label{eq:TypeIIpowerlaw}
a &\propto& e^{\tepsilon \tN},\cr
H &\propto& e^{- \tN},\cr
\gamma &\propto& e^{\ts \tN}.
\ea
We can recognize this solution as power-law contraction from the definition of the number of e-folds (\ref{eq:TypeIN}),
\be
dt \propto d \tN / H \propto - e^{\tN} d \tN,
\ee
with solution
\be
t \propto - e^{\tN},
\ee
where the minus sign indicates that $t$ is negative and tending toward zero for $d \tN < 0$. Then, from (\ref{eq:TypeIIpowerlaw}),
\ba
\label{eq:TypeIIt}
a(t) &\propto& (-t)^{\tepsilon},\cr
H(t) &\propto& t^{-1},\cr
\gamma(t) &\propto& (-t)^{\ts}.
\ea

As in the Type I case, we can fix the gauge to a DBI Lagrangian by taking ${\mathcal L}_X = \gamma$, so that we can define a dual flow hierarchy as:
\ba
\label{eq:TypeIIflowparams}
\tepsilon\left(\phi\right) &=& \frac{2 M_P^2}{\gamma\left(\phi\right)} \left(\frac{a'\left(\phi\right)}{a\left(\phi\right)}\right)^2,\cr
\ts\left(\phi\right) &=& \frac{2 M_P^2}{\gamma\left(\phi\right)} \left(\frac{\gamma'\left(\phi\right)}{\gamma\left(\phi\right)}\right) \left(\frac{a'\left(\phi\right)}{a\left(\phi\right)}\right),\cr
\teta\left(\phi\right) &\equiv& \frac{2 M_P^2}{\gamma\left(\phi\right)} \frac{a''\left(\phi\right)}{a\left(\phi\right)},\cr
\trho\left(\phi\right) &\equiv& \frac{2 M_P^2}{\gamma\left(\phi\right)}  \frac{\gamma''\left(\phi\right)}{\gamma\left(\phi\right)},\cr
{}^\ell \tlambda\left(\phi\right) &\equiv& \left(\frac{2 M_P^2}{\gamma\left(\phi\right)}\right)^{\ell} \left(\frac{a'\left(\phi\right)}{a\left(\phi\right)}\right)^{\ell - 1} \frac{1}{a\left(\phi\right)} \frac{d^{\ell + 1} a\left(\phi\right)}{d \phi^{\ell + 1}},\cr
{}^\ell \talpha\left(\phi\right) &\equiv& \left(\frac{2 M_P^2}{\gamma\left(\phi\right)}\right)^{\ell} \left(\frac{a'\left(\phi\right)}{a\left(\phi\right)}\right)^{\ell - 1} \frac{1}{\gamma\left(\phi\right)} \frac{d^{\ell + 1} \gamma\left(\phi\right)}{d \phi^{\ell + 1}}.
\ea
This is just the $a \leftrightarrow H$ dual of the Type I flow hierarchy (\ref{eq:TypeIflowparams}). Since $d \tN$ is also the $a \leftrightarrow H$ dual of $d N$ (\ref{eq:DBIN}),
\be
d \tN = \frac{\gamma}{2 M_P^2} \left(\frac{a}{a'}\right) d \phi,
\ee
it follows immediately that the flow equations (\ref{eq:TypeIflowequations}) are invariant under the duality,
\ba
\label{eq:TypeIIflowequations}
\tepsilon &=& \frac{1}{a}\frac{d a}{d \tN},\cr
\frac{d \tepsilon}{d \tN} &=& \tepsilon\left(2 \teta - 2 \tepsilon - \ts\right),\cr
\frac{d \teta}{d \tN} &=& -\teta\left(\tepsilon + \ts\right) + {}^2 \tlambda,\cr
\frac{d {}^\ell \tlambda}{d \tN} &=& -{}^\ell \tlambda \left[\ell \left(\ts + \tepsilon\right) - \left(\ell - 1\right) \teta\right] + {}^{\ell + 1}\tlambda,\cr
\ts &=& \frac{1}{\gamma}\frac{d \gamma}{d \tN},\cr
\frac{d \ts}{d \tN} &=& -\ts \left(2 \ts + \tepsilon - \teta\right) + \tepsilon \trho,\cr
\frac{d \trho}{d \tN} &=& -2 \trho \ts + {}^2 \talpha,\cr
\frac{d {}^\ell \talpha}{d \tN} &=& -{}^\ell \talpha \left[\left(\ell + 1\right) \ts + \left(\ell - 1\right) (\tepsilon-\teta)\right] + {}^{\ell + 1} \talpha.
\ea
This allows us to realize a dual of the slow roll limit in inflation, such that in the small-parameter limit $\tepsilon \ll 1$, $\ts \ll 1$, $\teta \ll 1$ $\ldots$ the flow parameters are approximately constant,
\ba
&&\frac{d \tepsilon}{d \tN} \simeq 0,\cr
&&\frac{d \ts}{d \tN} \simeq 0,\cr
&&\frac{d \teta}{d \tN}\simeq 0\ \ldots
\ea
From (\ref{eq:TypeIIt}), we see that the dual slow-roll limit $\tepsilon \ll 1$ and $\ts \ll 1$ corresponds to quasi-static $a$ and $\gamma$, with rapidly varying Hubble parameter, which can be identified as cyclic contraction \cite{Steinhardt:2001st,Khoury:2003rt}.

\subsection{Type III cosmology}

In this section, we construct a new family of cosmological solutions using the $a \leftrightarrow \gamma$ duality
\ba \label{defdbbn:TypeIII}
&&\frac{1}{\gamma}\frac{d \gamma}{d \bbN} \equiv -1,\cr
&&\bbepsilon \equiv \frac{1}{H}\frac{d H}{d \bbN},\cr
&&\bbs \equiv \frac{1}{a}\frac{d a}{d \bbN}.
\ea
The dual number of e-folds $\bbN$ can be related to the number of e-folds $N$ by:
\ba
\label{eq:TypeIIIdifferential}
d \bbN &=& - \frac{d \gamma}{\gamma} = \frac{1}{\bbepsilon} \frac{d H}{H} = \frac{1}{\bbs} \frac{da}{a}\cr
&=& - \frac{1}{\bbs} d N = \frac{1}{\bbs} H dt.
\ea
Whether $d \bbN < 0$ corresponds to an expanding or a contracting universe in this case depends on the sign of the parameter $\bbs$. Using (\ref{eq:TypeIIIdifferential}), we can relate the parameters $\bbepsilon$ and $\bbs$ to their Type I counterparts,
\ba
&&\bbepsilon = - \frac{\epsilon}{s},\cr
&&\bbs =  \frac{1}{s}.  
\ea

As in Type I and Type II cosmologies, we can relate the acoustic horizon size to the number of e-folds by
\be
d {\mathcal N} = \left(1 - \bbepsilon - \bbs\right) d \bbN.
\ee
We construct an exact solution for the spacetime evolution by taking $\bbepsilon = {\rm const.}$ and $\bbs = {\rm const.}$, 
\ba
\label{eq:TypeIIIpowerlaw}
a &\propto& e^{\bbs \bbN},\cr
H &\propto& e^{\bbepsilon \bbN},\cr
\gamma &\propto& e^{- \bbN}.
\ea
We can recognize this solution as power-law evolution from the definition of the number of e-folds (\ref{eq:TypeIIIdifferential}),
\be
dt \propto d \bbN / H \propto e^{-\bbepsilon \bbN} d \bbN,
\ee
with solution
\be
t \propto e^{-\bbepsilon \bbN}.
\ee
Then, from (\ref{eq:TypeIIIpowerlaw}),
\ba
\label{eq:TypeIIIt}
a(t) &\propto& t^{-\bbs / \bbepsilon},\cr
H(t) &\propto& t^{-1},\cr
\gamma(t) &\propto& t^{1/ \bbepsilon}.
\ea
Using (\ref{eq:TypeIIIdifferential}) and (\ref{eq:usefulidentity}) gives an expression for $\bbepsilon$ in terms of the field $\phi$,
\be
\bbepsilon = \frac{{\mathcal L}_X}{2 M_P^2} \left(\frac{a\left(\phi\right)}{a'\left(\phi\right)}\right) \left(\frac{\gamma\left(\phi\right)}{\gamma'\left(\phi\right)}\right).
\ee
Similarly,
\be
\bbs = \frac{{\mathcal L}_X}{2 M_P^2} \left(\frac{H\left(\phi\right)}{H'\left(\phi\right)}\right) \left(\frac{\gamma\left(\phi\right)}{\gamma'\left(\phi\right)}\right).
\ee
Note that these expressions are {\it not} simply the $a \leftrightarrow \gamma$ dual of the expressions (\ref{eq:TypeIepsilon}) and (\ref{eq:TypeIs}). However, note that the DBI gauge condition
\be
\label{eq:DBIgaugecondition}
{\mathcal L}_X = - 2 M_P^2 \left(\frac{H'}{H}\right) \left(\frac{a'}{a}\right) = \gamma
\ee
is invariant under the duality $a \leftrightarrow H$, so that both Type I and Type II models retain their DBI form under the duality. The DBI gauge condition (\ref{eq:DBIgaugecondition}), however, is not invariant under the duality $a \leftrightarrow \gamma$, which results in a gauge condition
\be
\label{eq:TypeIIIidentity}
\frac{2 M_P^2}{a} \left(\frac{H'}{H}\right) \left(\frac{\gamma'}{\gamma}\right) = -1,
\ee
which is the $a \leftrightarrow \gamma$ dual of the identity (\ref{eq:TypeIidentity}). 
The identity (\ref{eq:usefulidentity}) always holds, and multiplying by (\ref{eq:TypeIIIidentity}) we can equivalently write the Type III gauge condition as
\be
\label{eq:TypeIIIgauge}
{\mathcal L}_X = \frac{4 M_P^4}{a} \left(\frac{a'}{a}\right) \left(\frac{\gamma'}{\gamma}\right) \left(\frac{H'}{H}\right)^2.
\ee
The parameters $\bbepsilon$ and $\bbs$ are then given by
\ba
\label{eq:TypeIIIepsilons}
\bbepsilon &=& \frac{2 M_P^2}{a\left(\phi\right)} \left(\frac{H'\left(\phi\right)}{H\left(\phi\right)}\right)^2,\cr
\bbs &=&  \frac{2 M_P^2}{a\left(\phi\right)} \left(\frac{a'\left(\phi\right)}{a\left(\phi\right)}\right) \left(\frac{H'\left(\phi\right)}{H\left(\phi\right)}\right).
\ea
Therefore a single gauge choice (\ref{eq:TypeIIIgauge}) self-consistently defines both $\bbepsilon$ and $\bbs$ as the the $a \leftrightarrow \gamma$ dual of the Type I parameters (\ref{eq:DBIepsilons}). From Eqs. (\ref{eq:TypeIIIepsilons}) and (\ref{eq:TypeIIIidentity}), we can write the gauge condition (\ref{eq:TypeIIIgauge}) in the simple form
\be
\label{eq:TypeIIIgaugesimple}
{\mathcal L}_X = -a \bbs.
\ee 
We can then define a Type III flow hierarchy as the $a \leftrightarrow \gamma$ dual of the Type I flow hierarchy (\ref{eq:TypeIflowparams})
\ba
\label{eq:TypeIIIflowparams}
\bbeta\left(\phi\right) &\equiv& \frac{2 M_P^2}{a\left(\phi\right)} \frac{H''\left(\phi\right)}{H\left(\phi\right)},\cr
\bbrho\left(\phi\right) &\equiv& \frac{2 M_P^2}{a\left(\phi\right)}  \frac{a''\left(\phi\right)}{a\left(\phi\right)},\cr
{}^\ell \bblambda\left(\phi\right) &\equiv& \left(\frac{2 M_P^2}{a\left(\phi\right)}\right)^{\ell} \left(\frac{H'\left(\phi\right)}{H\left(\phi\right)}\right)^{\ell - 1} \frac{1}{H\left(\phi\right)} \frac{d^{\ell + 1} H\left(\phi\right)}{d \phi^{\ell + 1}},\cr
{}^\ell \bbalpha\left(\phi\right) &\equiv& \left(\frac{2 M_P^2}{a\left(\phi\right)}\right)^{\ell} \left(\frac{H'\left(\phi\right)}{H\left(\phi\right)}\right)^{\ell - 1} \frac{1}{a\left(\phi\right)} \frac{d^{\ell + 1} a\left(\phi\right)}{d \phi^{\ell + 1}}.
\ea
Using Eq. (\ref{eq:TypeIIIidentity}), we can express the number of e-folds in terms of the field $\phi$ as:
\be
d \bbN = - \frac{\gamma'}{\gamma} d \phi = - \frac{a}{2 M_P^2} \left(\frac{H}{H'}\right) d \phi,
\ee
which is the $a \leftrightarrow \gamma$ dual of the Type I expression (\ref{eq:DBIN}). Therefore the flow equations are again invariant,
\ba
\label{eq:TypeIIIflowequations}
\bbepsilon &=& \frac{1}{H}\frac{d H}{d \bbN},\cr
\frac{d \bbepsilon}{d \bbN} &=& \bbepsilon\left(2 \bbeta - 2 \bbepsilon - \bbs\right),\cr
\frac{d \bbeta}{d \bbN} &=& -\bbeta\left(\bbepsilon + \bbs\right) + {}^2 \bblambda,\cr
\frac{d {}^\ell \bblambda}{d \bbN} &=& -{}^\ell \bblambda \left[\ell \left(\bbs + \bbepsilon\right) - \left(\ell - 1\right) \bbeta\right] + {}^{\ell + 1}\bblambda,\cr
\bbs &=& \frac{1}{a}\frac{d a}{d \bbN},\cr
\frac{d \bbs}{d \bbN} &=& -\bbs \left(2 \bbs + \bbepsilon - \bbeta\right) + \bbepsilon \bbrho,\cr
\frac{d \bbrho}{d \bbN} &=& -2 \bbrho \bbs + {}^2 \bbalpha,\cr
\frac{d {}^\ell \bbalpha}{d \bbN} &=& -{}^\ell \bbalpha \left[\left(\ell + 1\right) \bbs + \left(\ell - 1\right) (\bbepsilon-\bbeta)\right] + {}^{\ell + 1} \bbalpha.
\ea
From (\ref{eq:TypeIIIt}), we see that the dual slow-roll limit $\bbepsilon \ll 1$ and $\bbs \ll 1$ corresponds to quasi-static $a$ and $H$, with rapidly varying sound speed. This is a novel class of cosmological solution. 

\subsection{Discussion}

We have defined three types of cosmological model starting with the case of DBI inflation and applying dualities which leave the sound horizon
\be
R_s = \left(\gamma a H\right)^{-1}
\ee
invariant. From DBI inflation, which we call ``Type I'', we construct a ``Type II'' cosmology via the duality $a \leftrightarrow H$, the analog of the canonical duality discussed in Refs. \cite{Boyle:2004gv,Piao:2004uq,Lidsey:2004xd,Kinney:2010qa}. This duality maps an expanding universe onto a contracting universe, and vice-versa. We construct a Type III cosmology from Type I via the duality $a \leftrightarrow \gamma$, which gives a new class of cosmological models with slow expansion and rapidly varying speed of sound. We define the slow roll parameters $\epsilon$ and $s$ in each class of cosmology as follows.

Type I:
\ba
&&\epsilon = \frac{1}{H} \frac{d H}{d N} = \frac{2 M_P^2}{\gamma\left(\phi\right)} \left(\frac{H'\left(\phi\right)}{H\left(\phi\right)}\right)^2,\cr 
&&s = \frac{1}{\gamma}\frac{d \gamma}{d N} = \frac{2 M_P^2}{\gamma\left(\phi\right)} \left(\frac{\gamma'\left(\phi\right)}{\gamma\left(\phi\right)}\right) \left(\frac{H'\left(\phi\right)}{H\left(\phi\right)}\right).
\ea

Type II:
\ba
&&\tepsilon = \frac{1}{a} \frac{d a}{d \tN} = \frac{2 M_P^2}{\gamma\left(\phi\right)} \left(\frac{a'\left(\phi\right)}{a\left(\phi\right)}\right)^2 = \frac{1}{\epsilon},\cr
&&\ts = \frac{1}{\gamma}\frac{d \gamma}{d \tN} = \frac{2 M_P^2}{\gamma\left(\phi\right)} \left(\frac{\gamma'\left(\phi\right)}{\gamma\left(\phi\right)}\right) \left(\frac{a'\left(\phi\right)}{a\left(\phi\right)}\right) = - \frac{s}{\epsilon}.
\ea

Type III:
\ba
&&\bbepsilon = \frac{1}{H} \frac{d H}{d \bbN} = \frac{2 M_P^2}{a\left(\phi\right)} \left(\frac{H'\left(\phi\right)}{H\left(\phi\right)}\right)^2 = -\frac{\epsilon}{s},\cr 
&&\bbs = \frac{1}{a}\frac{d a}{d \bbN} = \frac{2 M_P^2}{a\left(\phi\right)} \left(\frac{a'\left(\phi\right)}{a\left(\phi\right)}\right) \left(\frac{H'\left(\phi\right)}{H\left(\phi\right)}\right) = \frac{1}{s}.
\ea

These fundamental parameters form the basis of an infinite hierarchy of ``flow'' parameters which completely describe the evolution of the space time via a set of first-order flow equations. The flow equations, remarkably, are invariant under the dualities. Each of these cosmologies admits an exact solution in which $\epsilon$ and $s$ are constant, and the physical parameters evolve as a power-law in coordinate time $t$. It is important to note that in the power-law case, the dualities defined here are in a sense trivial, since they simply amount to the re-naming of the basic parameters $\epsilon$ and $s$. For example, we can take the power-law solution for Type I,
\ba
a(t) &\propto& t^{1 / \epsilon},\cr
H(t) &\propto& t^{-1},\cr
\gamma(t) &\propto& t^{- s / \epsilon}.
\ea
and replace $\epsilon \rightarrow 1 / \tepsilon$, $s \rightarrow - \ts / \tepsilon$ to recover the Type II power-law solution
\ba
a(t) &\propto& t^{\tepsilon},\cr
H(t) &\propto& t^{-1},\cr
\gamma(t) &\propto& t^{\ts}.
\ea
There is no physics in this, just a redefinition of parameters. There are also manifestly trivial dualities such as $H \leftrightarrow \gamma$ in a Type I cosmology, which simply switches the definitions $\epsilon \leftrightarrow s$, and produces no new solutions. Similar trivial dualities are $a \leftrightarrow \gamma$ in a Type II cosmology, and $a \leftrightarrow H$ in a Type III cosmology. 

New physics results when we consider the dual of the slow roll limit of inflation, $\epsilon \ll 1$, $s \ll 1$. We can express the log of the comoving acoustic horizon ${\mathcal N} \equiv -\ln{\left(\gamma a H\right)}$ as
\ba 
d {\mathcal N} &=& \left(1 - \epsilon - s\right) dN\cr
 &=& \left(1 - \tepsilon - \ts\right) d\tN\cr
 &=& \left(1 - \bbepsilon - \bbs\right) d \bbN,
\ea
so that a shrinking acoustic horizon $d {\mathcal N} < 0$ corresponds to $d N < 0$ in slow-roll inflation and both the Type II and Type III duals to slow roll. However, in Type II dual slow-roll limit, $d \tN < 0$ corresponds to a {\it contracting} cosmology, with quasi-static scale factor and sound speed, and rapidly varying Hubble parameter. The Type III dual-slow roll limit has quasi-static scale factor and Hubble parameter, and rapidly varying speed of sound. Type III can be either expanding or contracting for $d \bbN < 0$, depending on the sign of $\bbs$. These are all physically distinct solutions. We can constrain the sign of $\bbs$ by applying the null energy condition, since
\be
\rho + p = 2 X {\mathcal L}_X \geq 0
\ee
implies that ${\mathcal L}_X$ is non-negative. Therefore, the gauge condition (\ref{eq:TypeIIIgaugesimple}) implies that 
\be
\bbs \leq 0,
\ee
which corresponds to an expanding cosmology $da \geq 0$ for $d \bbN < 0$. We will adopt the null energy condition in our analysis of perturbations. 

In the Type I and Type II cases, the field Lagrangian is of the DBI form as a result of the gauge choice 
\be
{\mathcal L}_X = \gamma,
\ee
so that 
\be
\gamma^2 = {\mathcal L}_X^2 = 1 + 2 X \frac{{\mathcal L}_{XX}}{{\mathcal L}_X},
\ee
resulting in a differential equation for ${\mathcal L}$,
\be
2 X {\mathcal L}_{XX} + {\mathcal L}_X - {\mathcal L}_X^3 = 0,
\ee
with solution
\be
{\mathcal L} = f^{-1}\left(\phi\right) \sqrt{1 + f\left(\phi\right) X} - V\left(\phi\right).
\ee
The Type III case is not in general of the DBI form, since a consistent definition of the duality requires a different gauge choice,
\be
{\mathcal L}_X = -a \bbs.
\ee
In the case $\bbs = {\rm const.}$, we can obtain a differential equation for the Lagrangian using
\be
\gamma \propto {\mathcal L}_X^{-1/\bbs},
\ee
so that \cite{Bessada:2009ns}
\be
\label{eq:TypeIIILagrangian}
2 X {\mathcal L}_{XX} + {\mathcal L}_X - C {\mathcal L}_X^{n} = 0,
\ee
where $C$ is a constant, and 
\be
n = 1 + \frac{2}{\bbs}.
\ee
The general solution for $n = {\rm const.}$ is a hypergeometric function. Particular cases result in familiar forms: For example, $n = 3$ is a DBI Lagrangian, and $n = 1$ is a canonical Lagrangian \cite{Bessada:2009ns}. Reconstruction of the Lagrangian for time-dependent $\bbs$ can be accomplished using parametric reconstruction methods \cite{Bean:2008ga}. 

In the next section, we discuss perturbations in Type I, II, and III cosmologies.

\section{Perturbations}
\label{sec:perturbations}

The quadratic action for curvature perturbation $\zeta$ around a flat Friedmann-Robertson-Walker (FRW) background with a time dependent sound speed $c_s(\tau)$ can in general be written as \cite{Brandenberger:1993zc, Garriga:1999vw} :
\be \label{action1}
S_2=\frac{M_{pl}^2}{2}\int dx^3 d\tau~ z^2 \left [ \left (\frac{d\zeta}{d\tau}\right)^2- c_s(\tau)^2(\nabla\zeta)^2 \right ], 
\ee
where $\tau$ is the conformal time, and
\be
 z\equiv \frac{a\sqrt{\rho+p}}{c_s H} \label{z}.
\ee
Introducing the canonical normalized variable $v=M_p z \zeta$, the equation of motion for $v$ in Fourier space is given by
\be
v''_k+(c_s^2k^2-\frac{z''}{z})v_k=0. \label{mode}
\ee 
(Here a prime denotes derivative with respect to conformal time $d\tau \equiv dt/a$.) Quantizing and solving the mode equation for $v_k$, one can then calculate the power spectrum of curvature perturbations
\be
P^\zeta_k \equiv \frac{1}{2\pi^2}|\zeta_k|^2k^3=\frac{k^3}{2\pi^2}\left|\frac{v_k}{z}\right|^2.
\ee
In Secs \ref{sec:TypeI}-\ref{sec:TypeIII} we first write the exact mode equations in term of flow parameters for the three types of cosmological evolution discussed in Sec. \ref{sec:background}. We then solve the equations in the small parameter limit, dual to slow-roll limit in DBI inflation, and find the spectral indices in each type. 

\subsection{Perturbations in Type I models}
\label{sec:TypeI}
In this section we review the solutions for scalar perturbations in DBI inflation in the slow-roll limit. (For more details see \cite{Kinney:2007ag}). In this case $z$ is given by
\be
z=\sqrt{2}M_p a \gamma \sqrt{\epsilon}.
\ee
Therefore $z''/z$ in terms of flow parameters can be written as:
\be
\frac{z''}{z} = (aH)^2 F(\epsilon,s,\eta, \rho, {}^2 \lambda), 
\ee
where
\be
F = 2+2\epsilon-3\eta-\frac{3}{2}s-4\epsilon\eta+\frac{1}{2}\eta s+2\epsilon^2+\eta^2-\frac{3}{4}s^2+{}^2 \lambda +\frac{1}{2}\epsilon \rho. \label{F}
\ee
We work with a dimensionless parameter $y$ instead of $\tau$, defined as the wavenumber in units of the inverse of the acoustic horizon size:
\be \label{defy}
y\equiv \frac{k}{aH\gamma}.
\ee
To write the mode equation in terms of $y$, we use the differential relation between $y$ and $\tau$
\be
\frac{d}{d\tau}= -aH(1-\epsilon-s)y\frac{d}{dy},
\ee
and for the second derivative we have
\be
\frac{d^2}{d\tau^2}=(aH)^2[(1-\epsilon-s)^2y^2\frac{d^2}{dy^2}+G(\epsilon,s,\eta,\rho)y\frac{d}{dy}],
\ee
where
\be
G = -s+3s^2+2\epsilon^2+3s\epsilon -2\epsilon \eta-s\eta-\epsilon\rho. \label{G}
\ee
Then the mode equation for $v$ in Fourier space is
\be
(1-\epsilon-s)^2y^2\frac{d^2v_k}{dy^2}+Gy\frac{dv_k}{dy}+(y^2-F)v_k=0,
\ee
where $F$ and $G$ are purely functions of flow parameters given by Eqs. (\ref{F}) and (\ref{G}), respectively. Solving the above second order ODE requires two boundary conditions. The relevant boundary conditions correspond to canonical quantization of $v$ and selection of the initial vacuum state, which is chosen to be Bunch-Davies vacuum state. 

The Bunch-Davies vacuum is defined as usual by taking the short wavelength limit of the mode equation (\ref{mode}) when $c_s^2k^2 \gg z''/z$. For cosmologies with time-dependent speed of sound, the mode equation in the ultraviolet limit has a WKB solution of the form:
\be
v_k= \frac{N_k}{\sqrt{c_sk}}\exp \left(\pm ik\int^\tau c_s d\tau\right).  
\ee
Selecting Bunch-Davies as the vacuum state of fluctuations in this limit is equivalent to choosing the negative exponent. We can rewrite the condition for the Bunch-Davies vacuum in terms of $y$ using the relation  
\be
\label{eq:defy}
dy = -c_sk(1-\epsilon-s)d\tau.
\ee
Therefore, in the small wavelength limit we have, for $\epsilon$ and $s$ approximately constant,
\be
v_k = \frac{1}{\sqrt{c_sk}}e^{iy/(1-\epsilon-s)}.
\ee

The evolution of $v_k$ at all times is found by solving the mode equation. In the slow-roll limit $\epsilon \ll 1$, $s \ll 1$, the mode equation is reduced to
\be
(1-2\epsilon-2s)y^2\frac{d^2v_k}{dy^2}-sy\frac{dv_k}{dy}+(y^2-2-2\epsilon+3\eta+\frac{3}{2}s)v_k=0. 
\ee
Keeping only the terms linear in flow parameters, the solution is given by
\be
v_k(y) \rightarrow y^{\frac{1}{2}(1+s)}\left[c_1 H_\mu^{(1)}\left(\frac{y}{1-\epsilon-s}\right)+c_2 H_\mu^{(2)}\left(\frac{y}{1-\epsilon-s}\right)\right], \label{modesolI}
\ee
where
\be
\mu \simeq \frac{3}{2}+2\epsilon+s-\eta.
\ee
Taking the asymptotic form of (\ref{modesolI}) in the short wavelength limit, $y\rightarrow \infty$, and writing the power of $y$ in front of Hankel function in terms of $c_s$ we can impose the Bunch-Davies vacuum correctly. Using the definition of the flow parameter $s$,
\be
s = \frac{1}{\gamma}\frac{d\gamma}{dN} = \frac{1}{\gamma}\frac{d\gamma}{dy}y(1-\epsilon-s),
\ee
and assuming that
\be
\frac{s}{1-\epsilon-s} \simeq {\rm const.},
\ee
we can express the speed of sound as a function of $y$ for a given mode:
\be
c_s \propto y^{\frac{-s}{1-\epsilon-s}}.
\ee 
To linear order in flow parameters, the speed of sound is then
\be
c_s \propto y^{-s}.
\ee
Imposing the Bunch-Davies vacuum corresponds to setting $c_2=0$ in Eq. (\ref{modesolI}). The constant $c_1$ is fixed by the canonical commutation relation for the field $v$,
\be
\left[v(\mathbf{x}, \tau), \pi(\mathbf{x'},\tau)\right]=i\delta^3(\mathbf{x}-\mathbf{x'}),
\ee
which corresponds to a Wronskian condition for the Fourier mode $v_k$,
\be
v_k \frac{\partial v_k^*}{\partial \tau}-v_k^*\frac{\partial v_k}{\partial \tau}=i.
\ee
The fully normalized solution $v_k$ is then given by
\be
v_k(y) = \frac{1}{2}\sqrt{\frac{\pi}{c_sk}}\sqrt{\frac{y}{1-\epsilon-s}}H_\mu^{(1)}\left(\frac{y}{1-\epsilon-s}\right).
\ee

The power spectrum for the curvature perturbation is calculated by taking the long wavelength limit
\ba
P_k^\zeta &=& \frac{k^3}{2\pi^2}\left|\frac{v_k}{z}\right|^2_{y \rightarrow 0} \cr
 &=& \frac{2^{2\mu-4}}{M_p^2}\left(\frac{\Gamma(\mu)}{\Gamma(3/2)}\right)^2 \frac{1}{c_s \epsilon}(1-2\epsilon-2s)\left(\frac{H}{2\pi}\right)^2y^{3-2\mu}.
\ea
The spectral index is then calculated by differentiating the above equation with respect to $k$ for $c_s/aH = const.$,
\be
n_\zeta -1 = 
\left|\frac{d \ln P_{\zeta}}{d\ln k}\right|_{c_s/aH=const.} = 3-2\mu = -4\epsilon+2\eta-2s,
\ee
which is the well-known result that DBI inflation in the slow-roll limit generates nearly scale-invariant scalar perturbations.
 
\subsection{Perturbations in Type II models}
\label{sec:TypeII}
In this section we derive the mode equation of $v_k$ in Type II cosmologies, cyclic scenarios, and solve it in the small parameter limit dual to slow-roll. Here $z$ is given by
\be
z=\frac{\sqrt{2}M_pa}{c_s\sqrt{\tepsilon}},
\ee
and therefore $z''/z$ in terms of flow parameters is
\be
\frac{z''}{z} = \left(\frac{aH}{\tepsilon}\right)^2 \tF(\tepsilon,\ts,\teta,\trho,\tlambda), 
\ee
where
\be
\tF = -2\tepsilon-\frac{3}{2}\ts + \teta +\frac{3}{4}\ts^2 +6 \tepsilon^2 + 3\teta^2 + 9\ts \tepsilon -\frac{9}{2}\ts\teta +\frac{3}{2} \tepsilon \trho-6\tepsilon \teta - {}^2\tlambda \tepsilon.
\ee
Here the relation between $y$ and $\tau$ in terms of flow parameters becomes
\be
\frac{d}{d\tau}= \left(\frac{aH}{\tepsilon}\right)(1-\tepsilon-\ts)y\frac{d}{dy},
\ee
and for the second derivative we have
\be
\frac{d^2}{d\tau^2}=\left(\frac{aH}{\tepsilon}\right)^2\left[(1-\tepsilon-\ts)^2y^2\frac{d^2}{dy^2}+\tG(\tepsilon,\ts,\teta,\trho)y\frac{d}{dy}\right],
\ee
where
\be
\tG =2\tepsilon-2\teta-\tepsilon\trho+\ts\teta+2\ts^2. \label{tG}
\ee
Then the mode equation for $v$ in Fourier space is
\be
(1-\tepsilon-\ts)^2 y^2\frac{d^2v_k}{dy^2}+\tG y\frac{dv_k}{dy}+(\tepsilon^2 y^2-\tF)v_k=0.
\ee
Similar to the canonical case discussed in \cite{Kinney:2010qa}, in order to obtain a consistent mode equation which allows us to correctly impose the Bunch-Davies vacuum, we use a new parameter $x$ which is related to $y$ by
\be
\frac{d}{dy} = - \tepsilon \frac{d}{dx}.
\ee 
Therefore, the relation between $x$ and $\tau$ in Type II cosmology is the same as that between $y$ and $\tau$ in Type I,
\ba
d y &=& - c_s k(1-\epsilon-s) d\tau,  \cr
d x &=& - c_s k (1-\tepsilon -\ts) d \tau. 
\ea
Note that the variable $y$ is a characteristic of the acoustic horizon $R_s$, while $x$ is the dual of the definition of $y$ in Eq. (\ref{eq:defy}). We show below that $x$ is the relevant dynamical variable for the mode function, with the result that perturbations in Type II cosmology freeze out at length scales much shorter than the acoustic horizon, $x \simeq \tepsilon y \sim 1 \ll y$.

The second derivatives are related by:
\be
\frac{d^2}{dy^2} = \frac {\tepsilon(2\teta-2\tepsilon-\ts)}{y(1-\tepsilon-\ts)} \frac{d}{dx} + \tepsilon^2 \frac{d^2}{dx^2}.
\ee
Hence the mode equation in terms of $x$ is
\be
(\tepsilon y)^2 (1-\tepsilon -\ts)^2 \frac{d^2v_k}{dx^2} + (-\ts+2\tepsilon^2+3\ts^2-\ts\teta+3\tepsilon\ts-\tepsilon\trho-2\tepsilon\teta)(\tepsilon y) \frac{dv_k}{dx}+(\tepsilon^2 y^2 - \tF))v_k=0.
\ee
To have a well-formulated Bessel equation we need to write the variable $\tepsilon y $ in terms of $x$. Using the differential relation between the two parameters and expanding $x$ as an infinite series in flow parameters and integrating by parts gives
\ba
x &=& - \int \tepsilon dy = - \tepsilon y + \int y \frac{d\tepsilon}{dy}dy \cr
 &=& -\tepsilon y + \frac{\tepsilon (2\teta-2\tepsilon-\ts)}{1-\tepsilon-\ts}y + \mathcal{O}(\tepsilon ^3,\tepsilon ^2 \teta, ...).  
\ea
Therefore, to first order in flow parameters
\be
x \simeq - \tepsilon y,
\ee
and the mode equation in dual-slow roll limit is given by
\be 
(1-2\tepsilon-2\ts)x^2\frac{d^2v_k}{dx^2} - \ts x \frac{dv_k}{dx} + (x^2 +\frac{3}{2} \ts -\teta +2\tepsilon)v_k=0,
\ee
with solution 
\be
v_k \rightarrow x^{\frac{1}{2}(1+\ts)}\left[c_1 H_\tmu^{(1)}\left(\frac{-x}{1-\tepsilon-\ts}\right)+c_2 H_\tmu^{(2)}\left(\frac{-x}{1-\tepsilon-\ts}\right)\right], \label{modesolII}
\ee
where the order of Hankel function $\tmu$ is given by
\be
\tmu \simeq \frac{1}{2}-2\tepsilon+3\ts+\teta.
\ee
Note that since $x \simeq \tepsilon y$ is the argument of the resulting Bessel function, mode freezing occurs at wavelengths much smaller than the acoustic horizon, $x \sim 1 \ll y$.
Again selecting the Bunch-Davies vacuum and canonical normalization fixes the two integration constants and the fully normalized solution is given by
\be
v_k(y) = \frac{1}{2}\sqrt{\frac{\pi}{c_sk}}\sqrt{\tepsilon y}H_\tmu^{(1)}\left(\tepsilon y\right).
\ee
To first order in the flow parameters, the spectral index for curvature perturbations is
\be
n_\zeta -1 = \left|\frac{d \ln P_{\zeta}}{d\ln k}\right|_{c_s/aH=const.} = 3-2\tmu = 2 + 2\tepsilon - 3\ts - \teta.
\ee 
Therefore in the small parameter limit dual to inflationary slow-roll, the spectrum of $\zeta$ approaches the well-known result of a blue spectrum. 

\subsection{Perturbations in Type III models}
\label{sec:TypeIII}
In this section we derive the mode equation for curvature perturbations in type III cosmology and solve the equation in the small parameter limit dual to slow-roll. In this case $z$ is given by
\be
z = \frac{\sqrt{2}M_pa}{c_s}\sqrt{\frac{\bbepsilon}{\bbs}}
\ee
and $z''/z$ in terms of flow parameters is
\be
\frac{z''}{z} = \left(\frac{aH}{\bbs}\right)^2 \bbF(\bbepsilon,\bbs,\bbeta,\bbrho,{}^2\bblambda,{}^2\bbalpha), 
\ee
where
\ba
\bbF = 1 &-& \bbepsilon-6\bbs+2\frac{\bbepsilon\bbrho}{\bbs}+\frac{1}{4}\bbepsilon^2+\frac{15}{4}\bbs^2-\frac{1}{4}\bbeta^2+\frac{1}{2} {}^2\bblambda+\frac{5}{4}(\frac{\bbepsilon\bbrho}{\bbs})^2  \cr 
&-& \bbs\bbepsilon +\frac{5}{2}\bbs\bbeta-\frac{5}{2}\bbepsilon\bbrho-\frac{1}{2}\bbepsilon\bbeta-\frac{1}{2}\frac{ {}^2\bbalpha \bbepsilon}{\bbs}+\frac{1}{2}\frac{\bbepsilon^2\bbrho}{\bbs}-\frac{\bbepsilon\bbrho\bbeta}{\bbs}.
\ea
Notice that in order to have self-consistent flow hierarchy, $\bbepsilon \bbrho/\bbs$ should be first order in flow parameters, since
\be
\label{eq:sflow}
\frac{1}{\bbs}\frac{d\bbs}{d\bbN}=-\left(2\bbs+\bbepsilon-\bbeta\right)+\frac{\bbepsilon\bbrho}{\bbs}.
\ee 

First, we write the mode equation in terms of the dimensionless parameter, $y$. In this case $y$ and $\tau$ are related in terms of flow parameters as
\be
\frac{d}{d\tau}= \frac{aH}{\bbs}(1-\bbepsilon-\bbs)y\frac{d}{dy},
\ee
and for the second derivative we have
\be
\frac{d^2}{d\tau^2}=\left(\frac{aH}{\bbs}\right)^2\left[(1-\bbepsilon-\bbs)^2y^2\frac{d^2}{dy^2}+\bbG(\bbepsilon,\bbs,\bbeta,\bbrho)y\frac{d}{dy}\right],
\ee
where
\be
\bbG = 1+\bbs-\bbeta-\frac{\bbepsilon\bbrho}{\bbs}+\bbepsilon^2-\bbs\bbepsilon-\bbepsilon\bbeta+\frac{\bbepsilon^2\bbrho}{\bbs}.\label{bG}
\ee
Then the mode equation for $v$ in Fourier space is
\be
(1-\bbepsilon-\bbs)^2 y^2\frac{d^2v_k}{dy^2}+\bbG y\frac{dv_k}{dy}+(\bbs^2 y^2-\bbF)v_k=0.
\ee
Here again by a change of variable and introducing a new variable $q$ instead of $y$, we obtain a consistent mode equation which allows us to impose the Bunch-Davies vacuum
\be
\frac{d}{dy} = - \bbs \frac{d}{dq}.
\ee 
The relation between $q$ and $\tau$ in Type III cosmology is the same as that between $y$ and $\tau$ in Type I,
\ba
d y &=& - c_s k(1-\epsilon-s) d\tau,  \cr  
d q &=& - c_s k (1-\bbepsilon -\bbs) d \tau. 
\ea

The second derivatives are related by:
\be
\frac{d^2}{dy^2} = \frac {\bbs(2\bbs+\bbepsilon-\bbeta)-\bbepsilon\bbrho}{y(1-\bbepsilon-\bbs)} \frac{d}{dq} + \bbs^2 \frac{d^2}{dq^2}.
\ee
The mode equation in terms of $q$ is then
\be
(\bbs y)^2 (1-\bbepsilon -\bbs)^2 \frac{d^2v_k}{dq^2} + (1-\bbepsilon-\bbs+2\bbepsilon^2+2\bbs^2-\bbepsilon\bbrho-\bbs\bbeta+2\bbs\bbepsilon-2\bbeta\bbepsilon)(-\bbs y) \frac{dv_k}{dq}+(\bbs^2 y^2 - \bbF)v_k=0.
\ee
Similar to the argument in Sec. \ref{sec:TypeII}, we have 
\ba
q &=& -\int \bbs dy = -\bbs y + \int y \frac{d\bbs}{dy}dy \cr
 &=& -\bbs y + \frac{-\bbs (2\bbs+\bbepsilon-\bbeta)+\bbepsilon\bbrho}{1-\bbepsilon-\bbs}y + \mathcal{O}(\bbs^3,\bbs ^2 \bbepsilon,...).  
\ea
Therefore, to first order in flow parameters,
\be
q \simeq -\bbs y,
\ee
and the mode equation in dual-slow roll limit is given by
\be 
(1-2\bbepsilon-2\bbs)q^2\frac{d^2v_k}{dq^2} +(1-\bbepsilon- \bbs) q \frac{dv_k}{dq} + \left(q^2 -1+\bbepsilon+6\bbs-2\frac{\bbepsilon\bbrho}{\bbs}\right)v_k=0,
\ee
with solution 
\be
v_k \rightarrow q^{-\frac{1}{2}(\bbepsilon+\bbs)}\left[c_1 H_\bbmu^{(1)}\left(\frac{-q}{1-\bbepsilon-\bbs}\right)+c_2 H_\bbmu^{(2)}\left(\frac{-q}{1-\bbepsilon-\bbs}\right)\right], \label{modesolIII}
\ee
where the order of Hankel function $\bbmu$ is given by
\be
\bbmu \simeq 1+\frac{1}{2}\bbepsilon-2\bbs+\left(\frac{\bbepsilon\bbrho}{\bbs}\right).
\ee
As in the type II case, modes freeze out at wavelengths much shorter than the acoustic horizon $q \sim 1 \ll y$.

After imposing the boundary conditions, the fully normalized solution is given by
\be
v_k(y) = \frac{1}{2}\sqrt{\frac{\pi}{c_sk}} \sqrt{\bbs y} H_\bbmu^{(1)}\left(\bbs y\right),
\ee
and the spectral index for curvature perturbations is
\be
n_\zeta -1 = \left|\frac{d \ln P_{\zeta}}{d\ln k}\right|_{c_s/aH=const.} =  3-2\bbmu  
 = 1-\bbepsilon +4\bbs-2\left(\frac{\bbepsilon\bbrho}{\bbs}\right) \qquad \quad
\ee
Therefore the third family of solutions results in blue spectrum for scalar perturbations.

\section{Conclusions}
\label{sec:conclusions}

In this paper, we have used the inflationary flow formalism \cite{Kinney:2002qn} to study cosmological models dual to inflation, where the duality is constructed such that the comoving acoustic horizon size
\be
R_s = \frac{c_S}{a H} = \left(\gamma a H\right)^{-1}
\ee
is invariant. This is the non-canonical generalization of the duality studied in Refs. \cite{Boyle:2004gv,Piao:2004uq,Lidsey:2004xd,Kinney:2010qa}, such that the comoving Hubble length
\be
R_H = \left(a H\right)^{-1}
\ee
remains invariant under the duality $a \leftrightarrow H$. The $a \leftrightarrow H$ dual of inflationary expansion is identified as cyclic contraction, with $a \simeq {\rm const.}$ and $H < 0$ \cite{Boyle:2004gv}. In the non-canonical generalization, the acoustic horizon is invariant under the dualities 
\ba
&&a \leftrightarrow H,\cr
&&a \leftrightarrow \gamma,\cr
&&H \leftrightarrow \gamma.
\ea

We construct cosmologies dual to inflation as follows: Begin with the case of DBI inflation, which we call a {\it Type I} cosmology, with dynamics parameterized by the number of e-folds $N$ and the flow parameters $\epsilon$ and $s$, such that
\ba
&&a \propto e^{-N},\cr
&&\epsilon \equiv \frac{1}{H}\frac{d H}{d N},\cr
&&s \equiv \frac{1}{\gamma} \frac{d \gamma}{d N}.
\ea
The null energy condition $\rho + p \geq 0$ implies that $\epsilon$ is positive-definite, since
\be
\epsilon = \frac{3}{2} \left(\frac{p}{\rho} + 1\right) \geq 0.
\ee
The parameter $s$ can be either positive or negative. 
The limit of slow roll inflation is $\epsilon \ll 1$, $s \ll 1$. The acoustic horizon size behaves as
\be
d {\mathcal N} \equiv d \left(\ln{R_s}\right) = \left(1 - \epsilon - s \right) d N,
\ee
so that the acoustic horizon shrinks for $d N < 0$, which corresponds to an expanding universe, $d a > 0$. Dualities under interchange of $a$, $H$, and $\gamma$ allow us to construct new cosmological solutions. The duality $a \leftrightarrow H$ corresponds to the non-canonical generalization of cyclic contraction, which we refer to as {\it Type II}. We define dual flow parameters $\tepsilon$ and $\ts$ as:
\ba
&&H \propto e^{-\tN},\cr
&&\tepsilon \equiv \frac{1}{a}\frac{d a}{d \tN},\cr
&&\ts \equiv \frac{1}{\gamma} \frac{d \gamma}{d \tN}.
\ea
These parameters can be expressed in terms of their counterparts in Type I cosmology as
\ba
&&\tepsilon = \frac{1}{\epsilon},\cr
&&\ts = - \frac{s}{\epsilon}.
\ea
Since $\epsilon$ is positive definite, so is $\tepsilon$, and $\ts$ can have either sign. 
In terms of the dual flow parameters, the acoustic horizon evolves as
\be
d {\mathcal N} \equiv d \left(\ln{R_s}\right) = \left(1 - \tepsilon - \ts \right) d \tN,
\ee
and again the acoustic horizon shrinks in the small parameter limit for $d \tN < 0$. However, in the Type II case, this corresponds to contraction, since $\tepsilon$ is positive definite, and therefore
\be
\frac{da}{a} = \tepsilon d \tN < 0.
\ee
We define a novel cosmology, {\it Type III}, via the duality $a \leftrightarrow \gamma$, with dual flow parameters defined as
\ba
&&\gamma \propto e^{-\tN},\cr
&&\bbepsilon \equiv \frac{1}{H}\frac{d H}{d \bbN},\cr
&&\bbs \equiv \frac{1}{a} \frac{d a}{d \bbN}.
\ea
These parameters can be expressed in terms of their counterparts in Type I cosmology as
\ba
&&\bbepsilon = - \frac{\epsilon}{s},\cr
&&\bbs = \frac{1}{s}.
\ea
In terms of the dual flow parameters, the acoustic horizon evolves as
\be
d {\mathcal N} \equiv d \left(\ln{R_s}\right) = \left(1 - \bbepsilon - \bbs \right) d \bbN,
\ee
with a shrinking acoustic horizon corresponding to $d \bbN < 0$. In Type III cosmology, we choose the gauge condition such that
\be
\rho + p = - \dot\phi^2 a \bbs,
\ee
so that the null energy condition requires $\bbs \leq 0$, corresponding to an expanding cosmology,
\be
\frac{da}{a} = \bbs d \bbN > 0.
\ee
Note that the duality $H \leftrightarrow \gamma$ maps between Type II and Type III cosmologies, forming a closed set of solutions under the set of dualities preserving the dynamics of the acoustic horizon. The Type III case is of particular interest, because such cosmologies have not been previously studied in the literature.\footnote{A similar, but distinct, slowly expanding cosmology was studied in Refs. \cite{Piao:2010bi,Liu:2011ns}.}

We derive the analog of the inflationary flow hierarchy (\ref{eq:TypeIflowparams}) and flow equations (\ref{eq:TypeIflowequations}) and find that they are invariant under all three dualities. The result is that in each case, taking the small parameter limit $\epsilon \ll 1$, $s \ll 1$ corresponds to the case where the flow parameters are approximately constant, analogous to the slow roll approximation in inflation. We refer to this as the {\it dual-slow roll} limit. We calculate the spectrum of scalar perturbations in all three scenarios in the dual-slow roll limit, with the following results:

Type I:
\be
n_\zeta -1 = -4\epsilon+2\eta-2s,
\ee

Type II:
\be
n_\zeta -1 = 2 + 2\tepsilon - 3\ts - \teta,
\ee 

Type III:
\be
n_\zeta - 1 = 1 - \bbepsilon + 4 \bbs - 2\left(\frac{\bbepsilon \bbrho}{\bbs}\right).
\ee
 
Type I cosmologies (DBI inflation) naturally approach scale invariance $n_\zeta \simeq 1$ in the slow roll limit $\epsilon \ll 1$, $\eta \ll 1$, $s \ll 1$. Type II (cyclic) cosmologies approach a strongly blue spectrum $n_\zeta \simeq 3$ in the dual-slow roll limit (although it is interesting to note that the corresponding Bardeen potential $\Phi$ is nearly scale-invariant \cite{Boyle:2004gv,Kinney:2010qa,Piao:2010bi}). Type III cosmologies also generate a blue spectrum of scalar perturbations, $n_\zeta \simeq 2$. 

Type III models represent a new type of cosmology, consisting of nearly static expansion 
\be
\frac{1}{a} \frac{da}{d \bbN} \ll 1,
\ee
with rapidly decreasing speed of sound, 
\be
\frac{1}{c_s}\frac{d c_s}{d \bbN} = 1,
\ee
which we call a {\it stalled} cosmology. Stalled cosmologies are an example of models which generate superhorizon perturbations via a superluminal sound speed, similar to ``tachyacoustic'' cosmology \cite{ArmendarizPicon:2003ht,ArmendarizPicon:2006if,Piao:2006ja,Magueijo:2008pm,Bessada:2009ns}. Unlike tachyacoustic models, which are characterized by matter- or radiation-dominated background evolution, stalled models are characterized by slow expansion and rapidly varying sound speed. In both the tachyacoustic and the stalled cases, the speed of sound is large at initial times, and therefore non-Gaussianity is suppressed.

If we do not impose the null energy condition, the parameter $\bbs$ is allowed to be positive, with $\rho + p < 0$, and 
\be
\frac{da}{a} = \bbs d \bbN < 0,
\ee
which corresponds to phantom-dominated contraction. We note the interesting possibility that Type III models may be useful for describing a nonsingular bounce \cite{Alexander:2007zm,Xue:2011nw}, using evolution from $\bbs > 0$ to $\bbs < 0$ , with $\rho + p < 0$ in a phantom contracting phase, and later to $\rho + p > 0$ in the expanding phase. 

Unlike Type I and II cosmologies, which are straightforwardly seen to be described by DBI field Lagrangians, reconstruction of the Lagrangian for Type III models is more complex, since Eq. (\ref{eq:TypeIIILagrangian}) for the Lagrangian depends directly on the flow parameters. For general flow parameters, parametric reconstruction is necessary; see Ref. \cite{Bean:2008ga} for a complete treatment. A related issue is that of gauge: we have constructed solutions dual to DBI inflation, which represents the particular choice of gauge ${\mathcal L}_X = \gamma$ in Type I. It would be of interest to construct dualities similar to those considered here in the more general gauge-independent formalism described in Ref. \cite{Bean:2008ga}. This is the subject of continuing work. 

\section*{Acknowledgements}
This research is supported  in part by the National Science Foundation under grants NSF-PHY-0757693 and NSF-PHY-1066278. WHK thanks the Kavli Institute for Cosmological Physics, where part of this work was completed, for generous hospitality. 

\bibliographystyle{JHEP}
\bibliography{bibliography}

\end{document}